\documentclass[final]{svjour3}

\usepackage{booktabs}
\usepackage{graphicx}
\usepackage{rotating}
\usepackage{amssymb}
\usepackage{siunitx}
\usepackage{array}
\usepackage{mathptmx}
\usepackage[numbers,sort]{natbib}
\usepackage[colorlinks]{hyperref}
\usepackage{indentfirst} %首行缩进
\usepackage{enumitem,lipsum}
\usepackage{float} 
\newcolumntype{M}[1]{>{\centering\arraybackslash}m{#1}}
\DeclareSIUnit\sq{\ensuremath{\Box}}

\makeatletter

\journalname{arXiv}
\bibpunct{[}{]}{,}{n}{}{,}

\begin{document}

\title{Cryogenic microwave performance of silicon nitride and amorphous silicon deposited using low-temperature ICPCVD}

\author{Jiamin Sun$^{1,2}$ \and Shibo Shu$^{3,\dagger}$ \and Ye Chai$^{3,4}$ \and Lin Zhu$^{1}$ \and Lingmei Zhang$^{2}$ \and Yongping Li$^{3}$ \and Zhouhui Liu$^{3}$ \and Zhengwei Li$^{3, \ddagger}$ \and Wenhua Shi$^{5,\S}$ \and Yu Xu$^{3}$ \and Daikang Yan$^{3}$ \and Weijie Guo$^{6}$ \and Yiwen Wang$^{7}$ \and Congzhan Liu$^{3}$}

\institute{$^1$ Shandong Institute of Advanced Technology, China\\
$^2$ Shandong University, China\\
$^3$ Institute of High Energy Physics, Chinese Academy of Sciences, China\\
$^4$ Xiangtan University, China\\
$^5$ Suzhou Institute of Nano-Tech and Nano-Bionics, Chinese Academy of Sciences, Suzhou 215123, China\\
$^6$ International Quantum Academy, Shenzhen 518048, China\\
$^7$ Southwest Jiaotong University, China\\
$^\dagger$\email{shusb@ihep.ac.cn}\\
$^\ddagger$\email{lizw@ihep.ac.cn}\\
$^\S$\email{whshi2007@sinano.ac.cn}}

\maketitle

\begin{abstract}
Fabrication of dielectrics at low temperature is required for temperature-sensitive detectors. For superconducting detectors, such as transition edge sensors and kinetic inductance detectors, AlMn is widely studied due to its variable superconducting transition temperature at different baking temperatures. Experimentally only the highest baking temperature determines AlMn transition temperature, so we need to control the wafer temperature during the whole process. In general, the highest process temperature happens during dielectric fabrication. Here, we present the cryogenic microwave performance of Si$_{3}$N$_{4}$, SiN$_{x}$ and $\alpha$-Si using ICPCVD at low temperature of 75 $^{\circ}$C. The dielectric constant, internal quality factor and TLS properties are studied using Al parallel plate resonators.

.

\end{abstract}

\section{Introduction} 
 
For cosmic microwave background observations, a dual-polarization dual-color pixel design is usually needed to maximize the detection efficiency. Such pixel design involves fabrication of multiple layers of dielectrics and metals. In projects like Simons Observatory (SO)~\cite{r17}, Advanced ACTPol~\cite{r24} and AliCPT~\cite{r25}, AlMn transition edge sensor (TES) is used. The advantage of AlMn superconducting alloy is that its transition temperature (\textit{T}$_\mathrm{c}$) is variable and can be adjusted to a suitable value by changing the doping concentration of manganese and more conveniently, changing the baking temperature after film deposition~\cite{r26}. This is both an advantage and a challenge. It is necessary to control the wafer temperature during the whole fabrication process. Due to the complexity of the fabrication, it is difficult to deposit AlMn after the dielectric. In this case, we need to explore low-temperature fabricated dielectrics. The current state of the art technology from SO is depositing SiN$_{x}$ by low-temperature plasma enhanced chemical vapor deposition (PECVD) at 150 $^{\circ}$C, after the AlMn process~\cite{r17}. Compared with PECVD, inductively coupled plasma chemical vapor deposition (ICPCVD) has a higher plasma density, and can achieve high-quality film deposition at a lower temperature. This work is an attempt to explore the low-temperature dielectric technology by ICPCVD, which is part of the fabrication for future Ali CMB projects~\cite{r18,r19}.\par

We started from a low temperature ICPCVD process at 75 $^{\circ}$C for silicon nitride (Si$_{3}$N$_{4}$), SiN$_{x}$ and $\alpha$-Si films. To characterize the cryogenic material properties of dielectrics, lumped-element resonators with parallel-plate capacitors (PPCs) were measured with a filling factor of 1.

\section{Test device design}

\begin{figure}[H]
    \centering
    \includegraphics[width=0.7\textwidth]{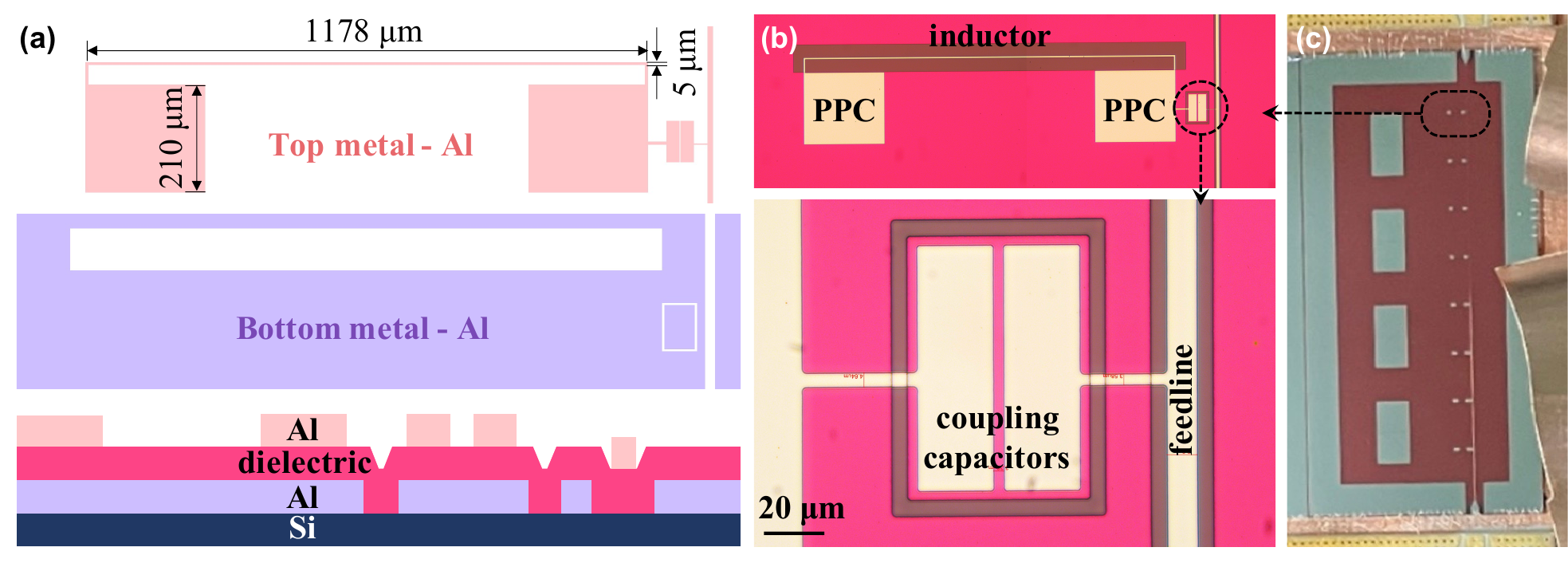}
    \caption{Design and fabrication of the test resonator device~\cite{r11}. (a) Layout and cross-section of the PPC lumped-element resonator. (b) Optical microscope picture of the as-fabricated device. (c) Photo of the sample mounted in a copper box and wire-bonded to two SMA connectors.}
    \label{fig:1}
\end{figure}

Each test resonator device consists of a lumped-element inductor and two PPCs~\cite{r11}, as shown in Fig.~\ref{fig:1}(a). The ground plane under the inductor is removed to minimize the capacitance in the inductor. The inductor has a length of $1178~\mathrm{\mu m}$, and a width of $5~\mathrm{\mu m}$ to minimize the resonance frequency variation resulted from lithographic accuracy. To minimize the frequency variation from kinetic inductance, we choose 200~nm-thick Al with a low kinetic inductance of 0.075~pH$/\ensuremath{\Box}$, giving a kinetic inductance ratio of $<2\%$. The PPCs use the same Al layer as top electrodes, and another 200~nm-thick Al layer as ground electrodes. This 300~nm thick dielectric layer was not patterned. The ground is connected through a big parallel capacitor on the edges of the chips. PPCs have a width of $210~\mathrm{\mu m}$. The length of PPCs varies from $230~\mathrm{\mu m}$ to $315~\mathrm{\mu m}$. Eight resonators are designed in two groups. A $50~\mathrm{\Omega}$ coplanar waveguide is used as the feedline, where the central trace is on the top layer and the gaps are in the ground plane. The coupling capacitors are also PPCs, directly connected to the feedline trace. 

\section{Device fabrication}

Device fabrication is done on a 100~mm-diameter high resistivity Si wafer ($>$10 k$\mathrm{\Omega}$$ \cdot$cm) to minimize possible loss from the substrate. Firstly, the bottom Al layer is deposited by electron beam evaporation (ULVAC ei-5z) at a rate of 0.2~nm/s, followed by UV lithography (SUSS MA6) and wet etching (standard phosphoric acid-based Al etchant). Next, a 300~nm-thick layer of dielectric film is deposited at low temperature using ICPCVD (Oxford PlasmaPro 100) with the parameters shown in Table \ref{tab:1}. Following the dielectric, the top Al layer is fabricated similarly to the bottom ones, which forms the Al capacitor pads and Al inductors of our resonators. The optical microscope pictures of the as-fabricated SiN$_{x}$ based PPC resonators are shown in Fig.~\ref{fig:1}(b). After dicing, the device is installed in a copper box for testing, as shown in Fig.~\ref{fig:1}(c).

\begin{table}[H]
\caption{Process parameters, refractive indices~($n$), and stresses of dielectric films.}
\centering
\renewcommand\arraystretch{1.2} %调整行间距
\begin{tabular}{ p{3cm}<{\centering} p{1.5cm}<{\centering} p{1.5cm}<{\centering} p{1.5cm}<{\centering}} %调整列宽及居中格式
 \toprule
Film & Si$_{3}$N$_{4}$ & SiN$_{x}$ & $\alpha$-Si\\
\hline % 一条横线\\
\midrule
SiH$_{4}$ (sccm) & 55 & 80 & 17\\
\hline
N$_{2}$ (sccm) & 160 & 50 & 0\\
\hline
Ar (sccm) & 0 & 0 & 80\\
\hline
ICP Power (W) & 1500 & 1500 & 1000\\
\hline
RF Power (W) & 3 & 0 & 100\\
\hline
Preasure (mT) & 20 & 20 & 10\\
\hline
Dep.Time (s) & 600 &4200 & 2550\\
\hline
T (℃) & 75 & 75 & 75\\
\hline
Thickness (nm) & 274 & 311 & 295\\
\hline
{\bf $n$$^*$ } & {\bf 1.88 } &{ \bf 2.85} & {\bf 3.05}\\ 
\hline
{\bf Stress (MPa)} & {\bf -82} & {\bf -84} & {\bf -228}\\ 
\bottomrule
\multicolumn{4}{l}{*measured using ellipsometer with a wavelength of 1600~nm at room temperature} \\

\label{tab:1}
\end{tabular}
\end{table}

Three different dielectrics, Si$_{3}$N$_{4}$, SiN$_{x}$ and $\alpha$-Si, are fabricated using ICPCVD at 75 $^{\circ}$C. Several process parameters including SiH$_{4}$/NH$_{3}$ flow ratio, ICP power, RF power, and process pressure are optimized for a low film stress. The optimal growth conditions and film performance are shown in Table \ref{tab:1}. 

RF power is one of the decisive factors affecting the film stress. Increasing the RF power during Si$_{3}$N$_{4}$ and SiN$_{x}$ deposition, changes the film stress from tensile stress (positive value) to compressive stress (negative value). This behavior could be explained by the inter-atomic extrusion caused by the Ar ions bombard under a high RF power.~\cite{r23} Here we tune the RF power to have the stress close to zero, as the dielectric will be on top of the narrow legs of the TES thermal island.

\section{Measurement results}

\subsection{Resonance frequency and dielectric constant}

\begin{figure}[H]
    \centering
    \includegraphics[width=1\linewidth]{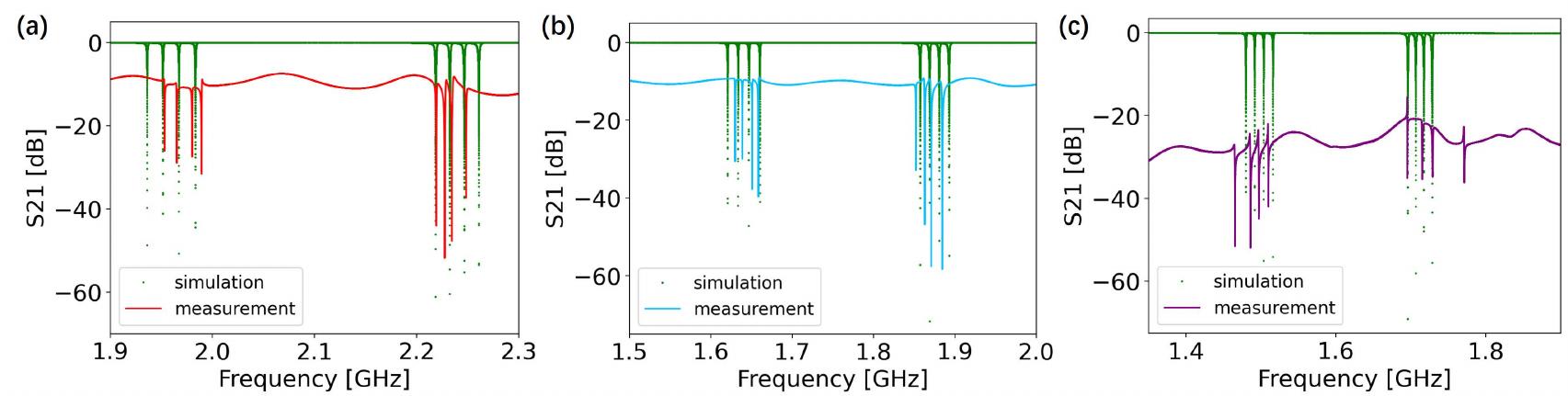}
    \caption{S21 of resonators with (a) Si$_{3}$N$_{4}$, (b) SiN$_{x}$ and (c) $\alpha$-Si dielectrics at 50 mK. The dielectric constants in simulation were manually adjusted to match the experimental data.}
    \label{fig:2}
\end{figure}

The S21 of three kinds of dielectrics are shown in Fig.~\ref{fig:2}. To estimate the dielectric constant, we ran electromagnetic simulation in Sonnet with different dielectric constants, and compared the simulation with measurements. The green dots in Fig.~\ref{fig:2} are simulated results with dielectric constants manually adjusted. The dielectric constants of Si$_{3}$N$_{4}$, SiN$_{x}$ and $\alpha$-Si are 7, 10, and 12, respectively. The fractional resonance frequency variation between measurement and simulation is smaller than $1\%$. This variation is mainly caused by the variation of dielectric thickness.

The amplitude difference between simulations and measurements of S21 in Fig.~\ref{fig:2} is due to the experimental setup, but the S21 of the $\alpha$-Si device has a -20~dB lower baseline than the other two devices suggesting a lossy on-chip transmission line. This S21 was very noisy and even lower at readout power smaller than -78~dBm, while the Si$_{3}$N$_{4}$ and SiN$_{x}$ curves were measured at -88~dBm. To explain this, we suggest that the $\alpha$-Si film has many pinholes~\cite{r20}, where was filled with Al. At low readout power, Al was superconducting and creates short points between the top and bottom metals. A high readout power turns those shorts points into normal metal, so the S21 and resonators becomes measurable but lossy, with a lower baseline of -20 dB than the other two. This result suggest that fabrication of $\alpha$-Si using ICPCVD at 75 $^{\circ}$C may not be capable.

\subsection{TLS loss}

Cryogenic dielectric loss can be described by a tunneling two-level system (TLS) model~\cite{r3}, which are found in most amorphous materials and arise from an energy difference between defect bond configurations coupled by tunneling, and generate an intrinsic excess noise at the metal/dielectric interface as well as in the bulk substrate that alters the resonance's quality factor and the sensitivity~\cite{r4,r5}.\par

\begin{figure}[H]
    \centering
    \includegraphics[width=1\linewidth]{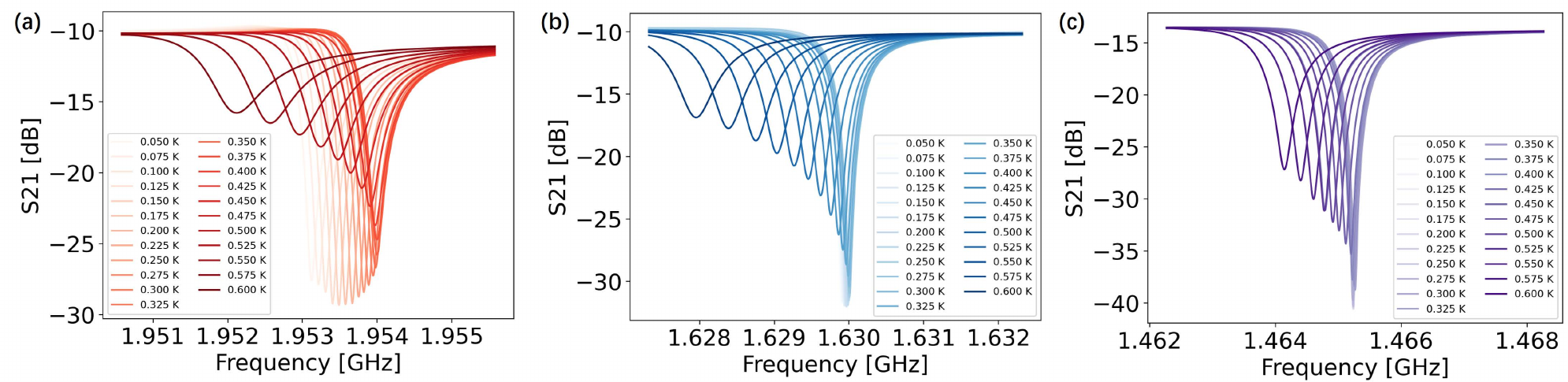}
    \caption{Resonance frequency of individual resonators with temperature sweeping from 50~mK to 600~mK. (a) Si$_{3}$N$_{4}$, (b) SiN$_{x}$ and (c) $\alpha$-Si dielectrics.}
    \label{fig:3}
\end{figure}

In a resonator, the dielectric modification by TLS effect can be derived assuming a log-uniform distribution of tunneling states~\cite{r12}. The subsequent shift in resonant frequency is given by:
 \begin{equation}
    \frac{f(T)-f_0}{f_0} = \frac{\mathrm{F}\delta_\mathrm{TLS}^0}{\pi}[\mathrm{Re\Psi}(\frac{1}{2}+\frac{1}{2\pi\mathrm{i}}\frac{\hbar\omega}{k_BT})-ln\frac{\hbar\omega}{k_BT}],
    \label{Eq:1}
\end{equation}
where \textit{f}($T$) is the resonator frequency at temperature \textit{T}, and \textit{f}$_{0}$ is the TLS-free resonance frequency. Filling factor F describes the fraction of the electric field energy that is contained in the TLS-hosting materials, $\delta$$_\mathrm{TLS}^0$ is the intrinsic TLS loss, $\mathrm \Psi$ is the complex digamma function, and \textit{k}$_{B}$ is the Boltzmann constant. Typically, F and $\delta$$_\mathrm{TLS}^0$ are degenerate when fitting \textit{f}($T$) vs. \textit{T} to the model, but we eliminated the degeneracy in our system by using PPCs, where F is close to one~\cite{r13}. The frequency shift of our individual resonators at lower temperatures can be described by the above-mentioned TLS model, as shown in Fig.~\ref{fig:3}. An increase in stage temperature can break Cooper pairs and change the kinetic inductance, resulting in a downshift of the resonance frequency~\cite{r14}, which is consistent with our measurement results.

\begin{figure}[H]
    \centering
    \includegraphics[width=1\linewidth]{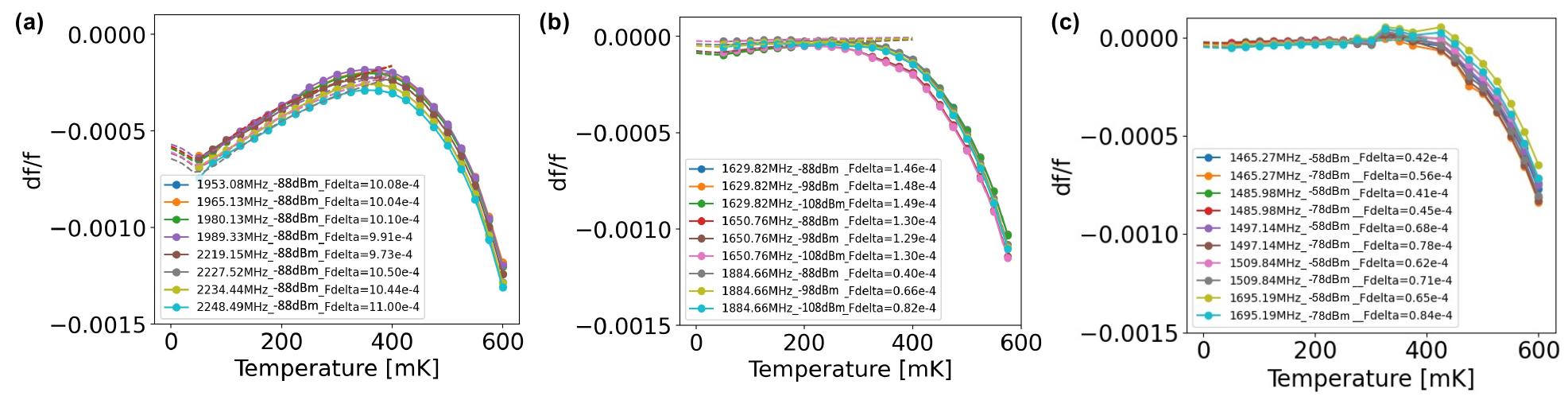}
    \caption{Resonance frequency shifts as a function of temperature. (a) Si$_{3}$N$_{4}$, (b) SiN$_{x}$, and (c) $\alpha$-Si dielectrics.} The dashed lines are fitting results of the frequency shift using Eq.~\ref{Eq:1}. The fitted TLS loss F$\delta$$_\mathrm{TLS}^0$ are labelled.
    \label{fig:4}
\end{figure}

We fitted the resonance frequency shift using Eq.~\ref{Eq:1}. It is quite obvious that TLS effect is more significant in Si$_{3}$N$_{4}$ than in SiN$_{x}$ from the level of frequency shift at the lowest temperature, shown in Fig.~\ref{fig:4}. The abnormal $\alpha$-Si data suggest that the fitting result does not represent the TLS effect. As filling factor F equals to one in PPCs, the fitted F$\delta$$_\mathrm{TLS}^0$ Table~\ref{tab:3} directly represent the TLS losses in Si$_{3}$N$_{4}$ and SiN$_{x}$ films. The TLS loss in SiN$_{x}$ is $4\sim 15 \times 10^{-5}$, one order of magnitude lower than $97\sim 110 \times 10^{-5}$ in Si$_{3}$N$_{4}$. Also, the high dielectric constant ${\varepsilon}_r=10$ of SiN$_{x}$ will also decrease the footprint size of PPCs in resonator designs.

\begin{table}[H]
\caption{Performances of PPC resonators with three kinds of dielectrics. The values with an asterisk (*) may not be unreliable.}
\centering
\renewcommand\arraystretch{1.2} %调整行间距
\begin{tabular}{ p{4cm}<{\centering} p{1.5cm}<{\centering} p{1.5cm}<{\centering} p{1.5cm}<{\centering}} %调整列宽及居中格式
 \toprule
Dielectric & Si$_{3}$N$_{4}$ & SiN$_{x}$ & $\alpha$-Si\\
\hline % 一条横线\\
\midrule
${\varepsilon}_r$ (ellipsometer @1600~nm) & 3.61 & 8.12 & 9.30\\
\hline
 ${\varepsilon}_r$ (resonator, GHz) & 7 & 10 & 12\\
 \hline
  F$\delta$$_\mathrm{TLS}^0$ ($ {\times} 10^{-5}$) &  97 $\sim$ 110 & 4 $\sim$ 15 & 4 $\sim$ 8$^*$\\
\hline
 $Q_\mathrm{i}$ ($\times 10^{5}$) & 0.1 $\sim$ 1.3 & 1.2 $\sim$ 2.0 & 0.2 $\sim$ 0.8$^*$\\
\hline
 %tan $\delta_\mathrm{i}$ = 
 1/$Q_\mathrm{i}$ ($\times 10^{-5}$) & 0.8 $\sim$ 8.0 & 0.5 $\sim$ 0.8 & 1.3 $\sim$ 5.0$^*$\\
\bottomrule
\label{tab:3}
\end{tabular}
\end{table}

\begin{figure}[H]
    \centering
    \includegraphics[width=1\linewidth]{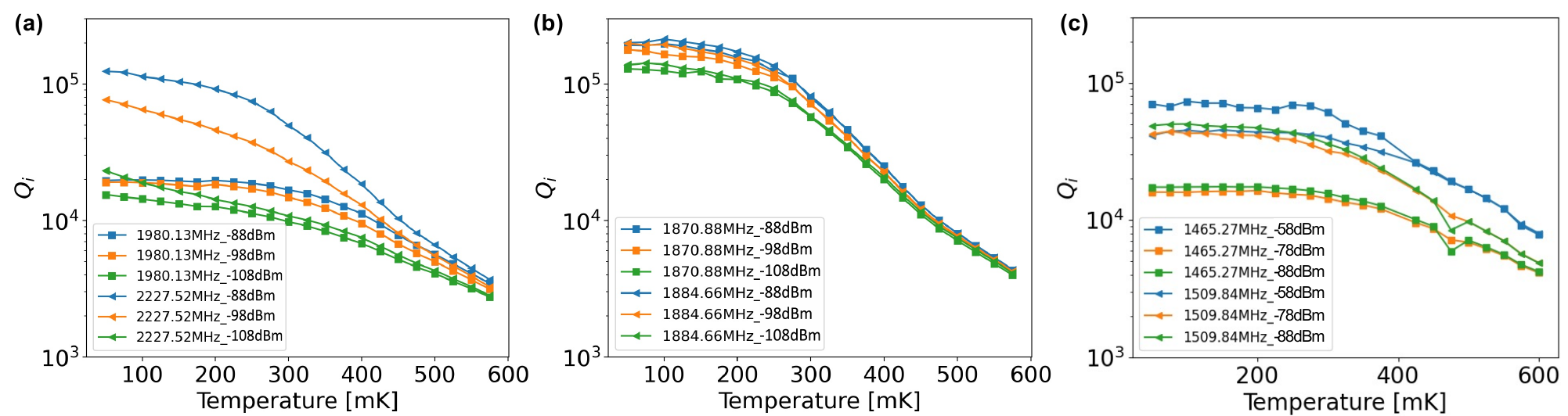}
    \caption{ Internal quality factors as a function of temperature. (a) Si$_{3}$N$_{4}$, (b) SiN$_{x}$ and (c) $\alpha$-Si.}
    \label{fig:5}
\end{figure}

We also studied the internal quality factor (\textit{Q}$_\mathrm{i}$) of these resonators. The \textit{Q}$_\mathrm{i}$ were measured at seven different powers from -58 dBm to -118 dBm. We find that the S21 curves for Si$_{3}$N$_{4}$ and SiN$_{x}$ are nonlinear at -58, -68 and -78 dBm, and noisy at -118 dBm. In this case, only S21 curves at -88 to -108 dBm for Si$_{3}$N$_{4}$ and SiN$_{x}$ are demonstrated. The suitable power for $\alpha$-Si is -58 to -78 dBm. As shown in Fig.~\ref{fig:5}, SiN$_{x}$ has a \textit{Q}$_\mathrm{i}$ of $1.2\sim 2.0 \times 10^{5}$, higher than Si$_{3}$N$_{4}$. Also the \textit{Q}$_\mathrm{i}$ of SiN$_{x}$ is less power dependent. As $\alpha$-Si based resonators behaved abnormally, the \textit{Q}$_\mathrm{i}$ data does not represent the film property. Compared with Si$_{3}$N$_{4}$, $\alpha$-Si seems to have higher \textit{Q}$_\mathrm{i}$, but it also has a much higher readout power. As the readout line has a lossy behavior, the readout power of $\alpha$-Si may not be compared with the other two films.

\subsection{TLS noise}

\begin{figure}[H]
    \centering
    \includegraphics[width=1\linewidth]{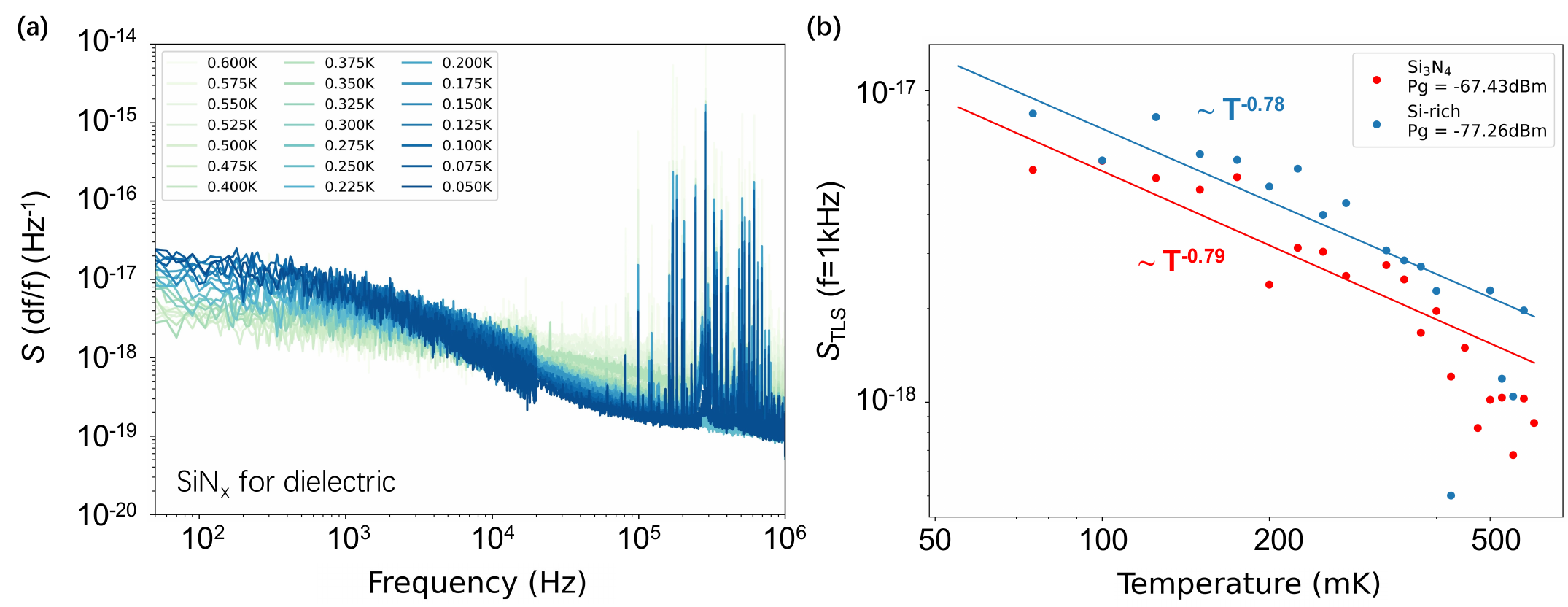}
    \caption{ TLS noise of resonators with (a) SiN$_{x}$ dielectric, (b) Si$_{3}$N$_{4}$ (red dots) and SiN$_{x}$ (blue dots) dielectrics.}
    \label{fig:6}
\end{figure}

The noise data were obtained using a homodyne detection scheme, which enables sampling of IQ data at 1~MHz. We first scan the S21 to obtain the resonant circle in the complex plane. The resonance point is where $\mid dz/df \mid$ has the maxima. Then the taken noise data is converted using the $dz/df$ vector at the resonance point.~\cite{r22}

The measured phase noise shows a typical $\sim$$f$$^{-0.5}$ TLS pattern. At 1~kHz, the TLS noise decreases with increasing temperature from 50~mK to 600~mK, shown in Fig.~\ref{fig:6}(b). The noise are $10\times 10^{-18}/$Hz and $3\times 10^{-18}/$Hz at 50~mK and 250~mK, respectively. The $\alpha$-Si resonators showed a much higher noise with a strong 50~Hz interference from mains utility power, so the data was not provided. The different readout power may be the reason for the higher noise value of SiN$_{x}$ than Si$_{3}$N$_{4}$ dielectric resonators. This result will benefit our kinetic inductance detector design and fabrication~\cite{r18}.

\section{Conclusion}

Cryogenic microwave performances of different dielectrics include Si$_{3}$N$_{4}$, SiN$_{x}$ and $\alpha$-Si deposited using low-temperature ICPCVD are studied in this work. The microwave dielectric constants of Si$_{3}$N$_{4}$, SiN$_{x}$ and $\alpha$-Si using PPCs at the range of $\sim$mK and $\sim$GHz are 7, 10, and 12, respectively.
In our specific fabrication conditions, SiN$_{x}$ shows a low TLS loss of $1\times 10^{-4}$, and a high internal quality factor of $\times 10^{5}$ with a unity filling factor. The TLS noise of SiN$_{x}$ is similar to Si$_{3}$N$_{4}$. At 250~mK, $S_{TLS}$ is around $3\times 10^{-18}/$Hz. The $\alpha$-Si deposited using this method may not be suitable for our application.\\

\noindent \textbf{Acknowledgements}

This work is supported by National Key Research and Development Program of China (Grant No. 2022YFC2205000), National Key Basic
Research and Development Program of China (Grant No. 2021YFC2203400), National Natural Science Foundation of China (Grant No. 2022YFC220013), Natural Science Foundation of Shandong Province (Grant No. ZR2022QA082), and China Postdoctoral Science Foundation (Grant No. 2021M700085).

\bibliography{Reference}
\bibliographystyle{JLTPv2}

\clearpage
\fontsize{15pt}{20pt}{\baselineskip 1cm}\selectfont

\noindent \textbf{Supplementary Material of \\}
\noindent \textbf{Cryogenic microwave performance of silicon nitride and amorphous silicon deposited using low-temperature ICPCVD}

\vspace{10mm}

\fontsize{10pt}{10pt}{\baselineskip 0cm}\selectfont
\noindent
\textbf{Table S1. Detailed calculations of the error between the simulation and measurement of dielectric constants.}
\begin{table}[htbp]
\centering
\renewcommand\arraystretch{1.5} %调整行间距
% \caption{\bf Detailed calculations of the error between the simulation and measurement of dielectric constants.}
\begin{tabular}{|c|c|c|c|c|c|c|c|c|c|}

\hline
Dielectrics & \multicolumn{3}{|c|}{Si$_{3}$N$_{4}$} & \multicolumn{3}{|c|}{SiN$_{x}$} & \multicolumn{3}{|c|}{$\alpha$-Si} \\

~ & \multicolumn{3}{|c|}{${\varepsilon}_{simulation} = 7$} & \multicolumn{3}{|c|}{${\varepsilon}_{simulation} = 10$} & \multicolumn{3}{|c|}{${\varepsilon}_{simulation} = 12$} \\

\hline
$L_{capacitor}$ & Measured & Simulated & Deviation & Measured & Simulated & Deviation & Measured & Simulated & Deviation\\

($\mu m$) & $f_\mathrm{res}$ (MHz) & $f_\mathrm{res}$ (MHz) & of $f_\mathrm{res}$ & $f_\mathrm{res}$ (MHz) & $f_\mathrm{res}$ (MHz) & of $f_\mathrm{res}$ & $f_\mathrm{res}$ (MHz) & $f_\mathrm{res}$ (MHz) & of $f_\mathrm{res}$ \\

\hline
230 & 1953.08 & 1935.96 & {0.88\%} & 1629.82 & 1620.59 & {0.57\%} & 1465.27 & 1479.58 & {0.97\%} \\

\hline
233 & 1965.13 & 1951.33 & 0.71\% & 1638.72 & 1633.62 & 0.31\% & 1485.98 & 1491.37 & 0.36\% \\

\hline
236 & 1980.13 & 1967.07 & 0.66\% & 1650.76 & 1646.80 & 0.24\% & 1497.14 & 1503.40 & 0.42\% \\

\hline
239 & 1989.33 & 1983.20 & 0.31\% & 1658.34 & 1660.52 & 0.13\% & 1509.84 & 1515.67 & 0.38\% \\

\hline
300 & 2219.15 & 2218.52 & 0.03\% & 1851.96 & 1857.55 & 0.30\% & 1695.19 & 1695.79 & 0.04\% \\

\hline
305 & 2227.52 & 2232.35 & 0.22\% & 1862.93 & 1869.03 & 0.33\% & 1715.35 & 1706.47 & 0.52\% \\

\hline
310 & 2234.44 & 2246.45 & 0.53\% & 1870.88 & 1880.54 & 0.51\% & 1728.97 & 1717.30 & 0.68\% \\

\hline
315 & 2248.49 & 2260.83 & 0.55\% & 1884.66 & 1892.60 & 0.42\% & 1771.18 & 1728.27 & 2.48\% \\

\hline
~ & \multicolumn{2}{|c|}{Average error} & 0.49\% & \multicolumn{2}{|c|}{Average error} & 0.35\% & \multicolumn{2}{|c|}{Average error} & 0.73\% \\

\hline
\end{tabular}
  \label{tab:shape-functions}
\end{table}

\noindent
\textbf{Table S2. Values of Fdelta with standard errors for lmfit.}
\begin{table}[htbp]
\centering
\renewcommand\arraystretch{1.5} %调整行间距
% \caption{\bf Values of Fdelta with standard errors for lmfit.}
\setlength{\tabcolsep}{5mm} {
\begin{tabular}{|c|c|c|c|c|c|}

\hline
\multicolumn{2}{|c|}{Si$_{3}$N$_{4}$} & \multicolumn{2}{|c|}{SiN$_{x}$} & \multicolumn{2}{|c|}{$\alpha$-Si} \\
\hline
Fdelta & Standard error for fit & Fdelta & Standard error for fit & Fdelta & Standard error for fit \\

\hline
1.01E-03 & 2.34E-05 & 1.46E-04 & 8.09E-06 & 4.16E-05 & 7.64E-06 \\
\hline
1.00E-03 & 4.54E-05 & 1.48E-04 & 8.01E-06 & 5.65E-05 & 1.74E-05 \\
\hline
1.01E-03 & 3.33E-05 & 1.49E-04 & 9.54E-06 & 4.13E-05 & 4.99E-06 \\
\hline
9.91E-04 & 2.25E-05 & 1.30E-04 & 1.17E-05 & 4.55E-05 & 1.73E-06 \\
\hline
9.73E-04 & 3.49E-05 & 1.29E-04 & 1.16E-05 & 6.77E-05 & 2.54E-06 \\
\hline
1.05E-03 & 2.43E-05 & 1.30E-04 & 1.33E-05 & 7.83E-05 & 2.75E-06 \\
\hline
1.04E-03 & 2.42E-05 & 3.98E-05 & 2.85E-06 & 6.19E-05 & 6.49E-06 \\
\hline
1.10E-03 & 5.90E-05 & 6.59E-05 & 8.96E-06 & 7.13E-05 & 1.12E-06 \\
\hline
~&~& 8.24E-05 & 1.58E-05 & 6.53E-05 & 1.26E-05 \\
\hline
~&~& ~&~& 8.37E-05 & 7.39E-06 \\

\hline
\end{tabular} }
  \label{tab:shape-functions}
\end{table}

\end{document}